\documentstyle[epsfig,prl,aps]{revtex}  

\newcommand{\tsub}[1]{_{\mbox{\scriptsize#1}}}

\newcommand{\tfrac}[2]{\mbox{$\small\frac{#1}{#2}$}}
\newcommand{\quarterthin}{\kern 0.0417em}
\newcommand{\comm}[2]{[ \quarterthin #1 , #2 \quarterthin ]}

\newcommand{\ket}[1]{|#1\rangle}
\newcommand{\ev}[1]{\langle#1\rangle}

\newcommand{\align}[1]{&#1&}

\begin{document}
\draft


 \twocolumn[\hsize\textwidth\columnwidth\hsize  
 \csname @twocolumnfalse\endcsname              

\title{
The SO(5) Theory as a
Critical Dynamical Symmetry}

\author{Lian-Ao Wu$^{(1)}$, Mike Guidry$^{(2)}$,
Yang Sun$^{(2)}$, and
Cheng-Li Wu$^{(3)}$
}
\address{
$^{(1)}$Department of Physics, Jilin University, Changchun, Jilin, 130023,
PRC\\
$^{(2)}$Department of Physics and Astronomy, University of Tennessee,
Knoxville, Tennessee 37996 \\
$^{(3)}$Department of Physics, Chung-Yuan Christian University, Taiwan, ROC}

\date{\today}
\maketitle

\begin{abstract}
We use generalized coherent states to analyze the $SO(5)$ theory of
high-temperature superconductivity and antiferromagnetism.  We demonstrate that
the $SO(5)$ symmetry
can be embedded in a larger algebra that
allows it to be
interpreted as a critical dynamical symmetry interpolating between
antiferromagnetic and superconducting phases.  This dynamical interpretation
suggests that $SO(5)$ defines a phase with the character of a spin-glass
for a significant range of doping.
\end{abstract}

\pacs{}

 ]  

\narrowtext

Data for
cuprate high-temperature superconductivity (SC) suggests
D-wave singlet pairing 
and that  SC in these systems
is closely related to the
antiferromagnetic (AF) insulator properties of
the undoped compounds.
Zhang et al
\cite{zha97,hen98,rab98}
proposed to unify AF and SC 
states by assembling their order parameters into a 5-dimensional vector
and constructing an $SO(5)$ group that rotates
AF order into D-wave SC order.
Recently \cite{gui99},
we introduced a $U(4)$ model of high-temperature
SC and AF order having $SO(5)$ as
a
subgroup.  In this paper we use coherent states
\cite{wmzha90} to analyze this subgroup and provide
a dynamical interpretation
of $SO(5)$ and the order parameters that it rotates.
The resulting energy surfaces suggest that
$SO(5)$ is a critical dynamical symmetry interpolating between AF and SC
phases that is extremely soft against
AF fluctuations.  Thus, we
shall propose that the Zhang $SO(5)$ theory corresponds to a spin-glass phase
that can serve as a doorway between AF and SC order.

Gilmore \cite{gil72,gil74} and Perelomov \cite{per72}
(see also earlier work by Klauder \cite{kla63}),
demonstrated that Glauber coherent states
\cite{gla63} for the
electromagnetic field could be generalized to coherent states for an arbitrary
Lie group.
These states can be analyzed in terms
of their geometry, which is in one-to-one correspondence
with the coset space.  However, it is often
simpler to view them as Hartree--Fock--Bogoliubov (HFB) variational
states constrained by the dynamical symmetry.
The symmetry-constrained HFB coherent
state method is discussed in Refs.\
\cite{wmzha90,wmzha87,wmzha88a,wmzha88,wmzha89}. 
It may be viewed as a type of mean-field approximation to
the underlying many-body problem that is particularly useful in the present
context
because it leads to easily visualized energy surfaces.
This provides a natural connection to spontaneously broken symmetries and
effective Lagrangian field theories. 

Let us begin with a group structure.
We introduce
16 bilinear fermion operators:
\begin{eqnarray}
p_{12}^\dagger&=&\sum_k g(k) c_{k\uparrow}^\dagger c_{-k\downarrow}^\dagger
\qquad
p_{12}=\sum_k g^*(k) c_{-k\downarrow} c_{k\uparrow} \nonumber
\\
q_{ij}^\dagger &=& \sum_k g(k) c_{k+Q,i}^\dagger c_{-k,j}^\dagger
\qquad q_{ij} = (q_{ij}^\dagger)^\dagger \label{eq1}
\\
Q_{ij} &=& \sum_k c_{k+Q,i}^\dagger c_{k,j} \qquad
S_{ij} = \sum_k c_{k,i}^\dagger c_{k,j}
- \tfrac12 \Omega \delta_{ij}  \nonumber
\end{eqnarray}
where $c_{k,i}^\dagger$ creates a fermion of momentum $k$ and spin
projection $i,j= 1 {\rm\ or\ }2 = \ \uparrow$ or
$\downarrow$, $Q=(\pi,\pi,\pi)$ is an AF
ordering vector, $\Omega$ is the lattice degeneracy, and
following Refs.\ \cite{hen98,rab98} we define $g(k) = {\rm sgn}
(\cos k_x -\cos k_y)$
with $g(k+Q) = -g(k)$ and
$\left| g(k) \right| = 1$.

Under commutation the operator set  (\ref{eq1}) closes a
$U(4)$ algebra  corresponding to the group structure
\begin{eqnarray}
&\supset& SO(4) \times U(1) \supset SU(2)\tsub{s} \times U(1) \nonumber
\\
U(4) \supset SU(4) &\supset& SO(5) \supset SU(2)\tsub{s} \times U(1)
\label{eq2}
\\
&\supset& SU(2)\tsub{p}
\times SU(2)\tsub{s} \supset SU(2)\tsub{s} \times U(1) 
\nonumber
\end{eqnarray}
where we require subgroup chains to end in
$SU(2)\tsub{s} \times U(1)$ representing spin and charge conservation.
In Ref.\ \cite{gui99} we discuss the representation structure
of (\ref{eq2}) and show that
the $SO(4)$ subgroup
is associated with AF and the $SU(2)\tsub{p}$
subgroup with
D-wave SC; in this paper,
we justify interpreting
the $SO(5)$ subgroup as a critical dynamical symmetry
leading to a  
``spin glass'' phase (SG).

It is convenient to
take as the generators of $U(4) \rightarrow U(1)\times SU(4)$ the
combinations
\begin{eqnarray}
Q_+&=&Q_{11}+Q_{22} = \sum_k (c_{k+Q\uparrow}^\dagger c_{k\uparrow}
+ c_{k+Q\downarrow}^\dagger c_{k\downarrow}) \nonumber
\\
\vec S &=& \left( \frac{S_{12}+S_{21}}{2},
                \ -i \, \frac {S_{12}-S_{21}}{2},
                \ \frac {S_{11}-S_{22}}{2} \right) \nonumber
\\
\vec {\cal Q} &=& \left(\frac{Q_{12}+Q_{21}}{2},\ -i\, \frac{Q_{12}-Q_{21}}{2},
\ \frac{Q_{11}-Q_{22}}{2} \right)
\label{eq3}
\\
\vec \pi^\dagger &=& \left(
i\frac {q_{11}^\dagger - q_{22}^\dagger}2, \
\frac{q_{11}^\dagger + q_{22}^\dagger}2,
\ -i\frac {q_{12}^\dagger + q_{21}^\dagger}2 \right) \nonumber
\\
\vec \pi&=&(\vec \pi^\dagger)^\dagger
\quad D^\dagger = p^\dagger_{12}
\quad D = p_{12}
\quad M=\tfrac12 (n-\Omega) \nonumber
\end{eqnarray}
where $Q_+$ generates the $U(1)$ factor and is associated
with charge density waves,
$\vec S$ is the spin operator, $\vec {\cal Q}$
is the staggered magnetization, the operators $\vec \pi^\dagger,
\vec \pi$ are those
of Ref.\  \cite{zha97}, the operators $D^\dagger, D$
are associated with
D-wave pairs, and $M$ is the charge operator.

To facilitate comparison with the $SO(5)$ symmetry the
$SO(6) \sim SU(4)$ generators may be expressed as
\begin{equation}
L_{ab} =
\left(
\begin{array}{cccccc}
0&&&&&
\\
D_+&0&&&&
\\
\pi_{x+} & -{\cal Q}_x & 0 &&&
\\
\pi_{y+} & -{\cal Q}_y & -S_z & 0 &&
\\
\pi_{z+} & -{\cal Q}_z & S_y & -S_x & 0 &
\\
i D_- & M & i\pi_{x-} & i \pi_{y-}
& i\pi_{z-} & 0
\end{array}
\right )
\label{eq4}
\end{equation}
\begin{equation}
D_\pm = \tfrac12 (D \pm D^\dagger)
\qquad
\pi_{i\pm} = \tfrac12 (\pi_i \pm  \pi^\dagger_i)
\label{eq5}
\end{equation}
with $L_{ab} = -L_{ba}$ and
with commutation relations
\begin{equation}
\comm{L_{ab}}{L_{cd}} = i(\delta_{ac} L_{bd}
-\delta_{ad}L_{bc} -\delta_{bc} L_{ad}+ \delta_{bd}L_{ac}).
\label{eq6}
\end{equation}

The coherent state method requires a faithful matrix representation of
$SU(4)$.  
Explicit multiplication verifies that the following
mapping preserves the algebra of  (\ref{eq6}):
\begin{eqnarray}
p_{12}^{\dagger }  \rightarrow \left[
\begin{array}{cc}
0 & i\sigma _y \\
0 & 0
\end{array}
\right] \qquad  p_{12}^{} \rightarrow \left[
\begin{array}{cc}
0 & 0 \\
-i\sigma _y & 0
\end{array}
\right]  \nonumber
\\
q_{12}^{\dagger }\rightarrow \left[
\begin{array}{cc}
0 & \sigma _x \\
0 & 0
\end{array}
\right] \qquad q_{12}^{}\rightarrow \left[
\begin{array}{cc}
0 & 0 \\
\sigma _x & 0
\end{array}
\right] \nonumber
\\
q_{11}^{\dagger }\rightarrow \left[
\begin{array}{cc}
0 & I+\sigma _z \\
0 & 0
\end{array}
\right] \qquad q_{11}^{}\rightarrow \left[
\begin{array}{cc}
0 & 0 \\
I+\sigma _z & 0
\end{array}
\right] \nonumber
\\
q_{22}^{\dagger }\rightarrow \left[
\begin{array}{cc}
0 & I-\sigma _z \\
0 & 0
\end{array}
\right] \qquad q_{22}^{}\rightarrow \left[
\begin{array}{cc}
0 & 0 \\
I-\sigma _z & 0
\end{array}
\right] \nonumber
\\
S_{12}^{}\rightarrow \left[
\begin{array}{cc}
\sigma _{+} & 0 \\
0 & -\sigma _{-}
\end{array}
\right] \qquad S_{21}^{}\rightarrow \left[
\begin{array}{cc}
\sigma _{-} & 0 \\
0 & -\sigma _{+}
\end{array}
\right]
\label{eq7}
\\
S_{11}-S_{22}\rightarrow \left[
\begin{array}{cc}
\sigma _z & 0 \\
0 & -\sigma _z
\end{array}
\right]
\qquad
Q_{12}^{}\rightarrow \left[
\begin{array}{cc}
\sigma _{+} & 0 \\
0 & \sigma _{-}
\end{array}
\right] \nonumber
\\
Q_{21}^{}\rightarrow \left[
\begin{array}{cc}
\sigma _{-} & 0 \\
0 & \sigma _{+}
\end{array}
\right]
\qquad
Q_{11}-Q_{22}\rightarrow \left[
\begin{array}{cc}
\sigma _z & 0 \\
0 & \sigma _z
\end{array}
\right] \nonumber
\\
N-\Omega =S_{11}+S_{22} \rightarrow \left[
\begin{array}{cc}
I & 0 \\
0 & -I
\end{array}
\right] \nonumber
\end{eqnarray}
where $\sigma _i$ are Pauli matrices, $\sigma
_{\pm }=\frac 12(\sigma _x\pm i\sigma _y)$ and $I$ is a unit
matrix.
The Casimir operator of
$SU(4) \simeq SO(6)$ is
\begin{equation}
C_{su(4)}=\vec \pi^\dagger \hspace{-2pt}\cdot \hspace{-2pt}\vec \pi
+ D^\dagger  D +
\vec S \hspace{-2pt}\cdot\hspace{-2pt} \vec S + \vec {\cal Q}
\hspace{-2pt}\cdot\hspace{-2pt} \vec {\cal Q} + M(M-4)
\label{eq8}
\end{equation}
and the irreps may be labeled by the $SO(6)$
quantum numbers, $(\sigma_1,\sigma_2,\sigma_3)$.
We take as a Hilbert space 
\begin{equation}
\ket{S} = \ket{n_x n_y n_z n_d} =
(\pi_x^\dagger)^{n_x}
(\pi_y^\dagger)^{n_y}
(\pi_z^\dagger)^{n_z}
(D^\dagger)^{n_d}
\ket{0}
\label{eq9}
\end{equation}
which is a collective subspace
associated with $SO(6)$ irreps of the form
$
(\sigma_1,\sigma_2,\sigma_3) = (\tfrac \Omega2,0,0).
$

Utilizing the methods of Ref.\ \cite{wmzha90}, the
coset space is $SU(4)/SO(4)\hspace{-2pt}\times\hspace{-2pt} U(1)$, where the
$SO(4)$ subgroup is generated by $\vec{\cal Q}$ and
$\vec S$ and $U(1)$ is generated by the charge operator $M$.
The coherent state may be written as
\begin{equation}
\mid \psi\rangle=\exp (\eta_{00} p_{12}^{\dagger }+\eta_{10} q_{12}^{\dagger
}-{\rm h.\ c.})\mid 0\rangle  \equiv {\cal T}\mid 0\rangle, 
\label{eq10}
\end{equation}
where the real parameters $\eta_{00}$ and $\eta_{10}$
are symmetry-constrained variational parameters.
Since they weight
the elementary excitation operators
$p^\dagger_{12}$ and $q^\dagger_{12}$ in Eq.\ (\ref{eq10}), they
represent collective state parameters for
a subspace truncated under
the $SU(4)$ symmetry.
The most general coherent state corresponds to a
4-dimensional, complex, compact manifold parameterized by 8 real variables.
The reduction of the coherent state
parameters to only two in Eq.\ (\ref{eq10})
follows from requiring spin and
time reversal symmetries for the most general wavefunction.

From the coset representative expressed in the 4-dimensional matrix
representation
(\ref{eq7}), the transformation ${\cal T}$ can be written as  
\begin{equation}
{\cal T} = \left[
\begin{array}{cc}
{\bf Y}_1 & {\bf X} \\ -{\bf X}^{\dagger } & {\bf Y}_2
\end{array}
\right]
\quad
{\bf X} \equiv \left[
\begin{array}{cc}
0 & \alpha + \beta \\
- (\alpha - \beta) & 0
\end{array}
\right]
\label{eq11}
\end{equation}
where ${\bf Y}_1$ and ${\bf Y}_2$ are
determined by the requirement that ${\cal T}$ be
unitary and $\alpha$ and $\beta$ are variational parameters 
related to $\eta_{00}$ and $\eta_{10}$ 
in Eq.\ (\ref{eq10}) (see Ref.\ \cite{wmzha90}).
Introducing $v_+ \equiv \alpha+\beta$ and  $v_- \equiv \alpha-\beta$,
and $u_+ = (1-v_+^2)^{1/2}$ and $u_- = (1-v_-^2)^{1/2}$,
the matrices can be written as:
$$
{\bf X} = \left[
\begin{array}{cc}
0 & v_+\\ -v_- & 0
\end{array}
\right]
\quad
{\bf Y}_1 = \left[
\begin{array}{cc}
u_+ & 0\\ 0 & u_-
\end{array}
\right]
\quad
{\bf Y}_2 = \left[
\begin{array}{cc}
u_- & 0\\ 0 & u_+
\end{array}
\right].
\label{XY-matrix}
$$
We may also introduce the gap parameters $\Delta_+ \equiv u_+v_+$ and
$\Delta_- \equiv u_-v_-$.

One can regard Eq.\ (\ref{eq10}) as a Bogoliubov type transformation,
which, through the operator $\cal T$, transforms the bare vacuum
state $|0\rangle$
to a
quasiparticle vacuum state $|\psi\rangle$.
Likewise, the basic fermion operators
$\{c^\dagger, c\}$ can be transformed to quasifermion operators
$\{a^\dagger, a\}$
through
\begin{equation}
{\cal T}
\left\{
\begin{array}{c}
c \\ c^\dagger
\end{array}
\right\}
{\cal T}^{-1}
\rightarrow \left\{
\begin{array}{c}
a \\ a^\dagger
\end{array}
\right\}.
\label{transf}
\end{equation}
Using the transformation (\ref{transf}), one can express
any one-body operator in the quasiparticle space \cite{ring80}.
In this way, many results known in HFB theory
can be applied directly in the present case.
Generally,
\begin{eqnarray}
\hat O &=& \sum_i {\cal O}^{(22)}_{i,i} + \sum_{i,j} [
({\cal O}^{(11)}_{i,j} - {\cal O}^{(22)}_{j,i})
a^\dagger_i a_j
\nonumber
\\
&&
+ {\cal O}^{(12)}_{i,j} a^\dagger_i a^\dagger_j
+ {\cal O}^{(21)}_{i,j} a_i a_j ], 
\label{quasi}
\end{eqnarray}
with the ${\cal O}$'s determined by the transformation properties
of the operator $\hat O$ (see 
Appendix E of Ref.\ \cite{ring80}).

The expectation value for an operator
$\hat O$ in the coherent state representation
is given by
$\langle \hat O\rangle = {\rm Tr} ({\cal O}^{(22)}) $ and
for a 2-body operator $\hat O^\dagger \hat O$,
\begin{eqnarray}
\langle\hat O^\dagger \hat O\rangle
\align= {\rm Tr} ({\cal O}^{\dagger(22)}) {\rm Tr} ({\cal O}^{(22)}) 
+ {\rm Tr} ({\cal O}^{\dagger(21)} {\cal O}^{(12)}).
\label{2body}
\end{eqnarray}
Utilizing (\ref{eq7}) and (\ref{eq11}),
the expectation value of the
particle number in the coherent state is
\begin{equation}
n\equiv \langle \hat N\rangle =2\Omega (\alpha ^2+\beta ^2)
= \Omega (v_+^2 + v_-^2).
\label{eq13a}
\end{equation}
From Eq.\ (\ref{eq13a}),
the squares of
$\alpha$ and $\beta$ are constrained by
the equation
of a circle having radius $\sqrt{n/2\Omega}$.
Thus, for fixed particle number $n$ we may evaluate matrix elements in terms of
a single variational parameter, say $\beta$,
which may be related to standard order parameters by comparing matrix elements.
For example, 
the $z$ component of the staggered magnetization is given by 
\begin{equation}
\langle {\cal Q}_z\rangle = 2\Omega \beta (n / (2\Omega) - \beta^2)^{1/2} 
= \tfrac12 \Omega (v_+^2 - v_-^2).
\label{eq14}
\end{equation}

Let us consider a simple $SO(5)$ Hamiltonian given by
$H_5 \sim C_{so(5)}$ where
the Casimir for $SO(5)$ is
\begin{equation}
C_{so(5)} =\vec \pi^\dagger \cdot \vec \pi + \vec S \cdot \vec S + M(M-3).
\label{eq16}
\end{equation}
We evaluate the $SO(5)$ ground-state energy surface by taking the expectation
value of $H_5$ in the coherent state (\ref{eq10}).
The expectation value of the $SO(5)$ Casimir operator may be written
as [see Eq.\ (\ref{eq8})]
\begin{equation}
\langle C_5\rangle =\ev{C_{su(4)}} -\langle D^{\dagger }D\rangle
-\langle \overrightarrow{\cal Q}\cdot \overrightarrow{\cal Q}\rangle 
+(n-\Omega )/2
\label{eq17}
\end{equation}
where $\ev{C_{su(4)}}=\frac \Omega 2(\frac \Omega
2+4)$.
From (\ref{eq7}), (\ref{eq11}), and (\ref{2body}),
\begin{eqnarray}
\langle D^{\dagger }D\rangle &=&\tfrac14 \Omega^2 (\Delta_+ + \Delta_-)^2
+ \tfrac12 \Omega (v_+^4 + v_-^4),
\label{eq18}
\\
\langle \overrightarrow{{\cal Q}}
\cdot \overrightarrow{{\cal Q}}\rangle &=&\tfrac14 \Omega^2 (v_+^2 - v_-^2)^2
\nonumber\\ 
& &+ \tfrac12 \Omega [\Delta_+^2 + \Delta_-^2 + (u_+v_- + u_-v_+)^2],
\label{eq19}
\end{eqnarray}
and the energy surface follows  from
Eqns.\ (\ref{eq16})--(\ref{eq19}).

Eq.\ (\ref{eq17}) is an expectation value of a Hamiltonian that can
be parameterized more generally as
\begin{equation}
H = -G_0 [(1-p)D^{\dagger }D +
p\overrightarrow{{\cal Q}}\cdot \overrightarrow{{\cal Q}}],
\label{eq20}
\end{equation}
with $G_0 > 0$ and $0 \leq p \leq 1$. 
Then $p = {1\over 2}$ corresponds to
$SO(5)$ symmetry, while the extreme values 0 and 1
correspond to  $SU(2)$ and $SO(4)$ symmetries, respectively (see Ref.\
\cite{gui99}).
Other values of $p$ respect $SU(4)$ symmetry but break the 
$SO(5)$, $SO(4)$, and $SU(2)$ subgroups.  
In Fig.\ 1 we illustrate the classical ground-state energy $\langle H \rangle$
as a function of the  order parameter $\beta$
for different 
lattice occupation fractions with $p = 0, {1\over 2}$ and 1.

For $p = 0$ [$SU(2)$ limit], the minimum energy occurs
at $\langle\beta\rangle = 0$ for all values of $n$, indicating
SC order.
For $p = 1$ [$SO(4)$ limit], the opposite situation occurs:
$\langle\beta\rangle = 0$ is an unstable point and an infinitessimal
fluctuation
will drive the system to the energy minima at
finite $\langle\beta\rangle$ (and thus finite $\langle{\cal Q}_z\rangle$), 
causing a transition to AF order.
From Fig.\ 1b, the
$SO(5)$ dynamical symmetry is seen to have extremely interesting
behavior:
For $n$ near $\Omega$ (half filling),
the system has an energy surface almost flat for
broad
ranges of $\beta$,
corresponding to large-amplitude fluctuations in
AF order.
We conclude that
the $SO(5)$ symmetry represents a
phase having much of the character of a spin glass
for a range of doping fractions.

Under exact $SO(5)$ symmetry, AF and SC states are degenerate at
half filling and there is no classical barrier between them (the
$n/\Omega =1$ curve of Fig.\ 1b).
To see clearly how the $SO(5)$ interpolates
between AF and SC states as particle number varies,
let us perturb slightly away from the $SO(5)$ limit of $p = 1/2$.
In Fig.\ 2a, results for $p = 0.52$ are shown.
We denote the value of $\langle\beta\rangle$
minimizing $\ev H$ as $\langle\beta\rangle_0$.
For small values of $n$ the stable point is
$\langle\beta\rangle_0 = 0$.
This corresponds to SC
order.
If $n$ is near $\Omega$ (half filling),
$\langle\beta\rangle_0 \simeq \pm 0.5$;
this
corresponds to AF order.
For intermediate values of $n$ there
is a
substantial region
in which the system has an energy surface almost flat for
broad
ranges of $\langle\beta\rangle$,
implying the presence of large-amplitude fluctuations in
AF order.  The corresponding variation of the AF order parameter
$\langle{\cal Q}_z\rangle$ with $n$ for several values of $p$ is summarized in
Fig.\ 2b.

Thus, as particle
number varies the $SO(5)$ symmetry, or slightly perturbed $SO(5)$ symmetry,
interpolates between AF order at half filling and
SC order at smaller filling, with the intermediate
regime acting like a spin glass.
Because for exact $SO(5)$ symmetry there is no barrier at half filling
between AF and SC states, one can fluctuate into the other at zero
cost in energy.  Only when $SO(5)$ is broken does the energy surface
interpolate between AF and SC order as
doping is varied (compare the surfaces for $p=1/2$ and $p=0.52$).
From Ref.\ \cite{gui99} 
and the present paper we conclude
that the underdoped portion of the 
cuprate phase diagram  is described by a Hamiltonian that
conserves 
$SU(4)$ but breaks (explicitly and weakly)
$SO(5)$ symmetry in a direction favoring AF order over SC
order.  
Thus, we propose that
the underdoped regime 
is associated with a conserved $SU(4)$ but 
weakly broken $SO(5)$ symmetry.
Likewise, optimally doped superconductors and AF insulators
are associated with the symmetries 
$SU(4) \supset SU(2)$ and  
$SU(4) \supset SO(4)$ respectively, or small perturbations around them.  

Dynamical symmetries that
interpolate between other dynamical
symmetries have been
termed  {\em critical dynamical symmetries} \cite{wmzha87}.  Thus, $SO(5)$ is
a critical dynamical symmetry.
Such symmetries are well known in nuclear structure
\cite{wmzha87,wmzha88a,wmzha90}, and the $SO(5)$ critical dynamical symmetry
discussed
here has many formal similarities with the $SO(7)$ critical dynamical symmetry
of the
(nuclear)
Fermion Dynamical Symmetry Model \cite{wu94}.
The condition for realization of the
$SO(5)$ phase is that the strength of
${\cal Q}\cdot {\cal Q}$ equals that of $D^{\dag} D$ in the Hamiltonian;
This is similar to 
the $SO(7)$ nuclear critical dynamical symmetry, which is realized when
there is an overall $SO(8)$ symmetry and the pairing and quadrupole interaction
terms are
equal in the Hamiltonian \cite{wmzha87}.

Finally, let us revisit a critical conclusion of Ref.\ \cite{gui99} in light of
the present
discussion.  In the
$SU(4)$ model, the $SO(5)$ subgroup defines only one
possible dynamical symmetry.  The AF phase near half filling
and the SC phase near optimal doping  of the $SU(4)$ model are
associated,
not with the $SO(5)$ subgroup, but with $SU(2)$ and $SO(4)$ subgroups,
respectively.
The $SO(5)$ subgroup
(identical to the Zhang $SO(5)$ but embedded in
a larger group structure) defines a phase interpolating between the $SU(2)$ and
$SO(4)$ phases
and characterized by large AF and SC fluctuations.
Thus, in the $SU(4)$ model the relationship of the AF phase near half filling
and the SC phase near optimal doping is {\em not constrained}
by the $SO(5)$ subgroup, 
and $SO(5)$ should be viewed as the appropriate symmetry for the
underdoped compounds.  

To conclude, we have proposed separately
\cite{gui99} that high $T\tsub c$ behavior
of the cuprates results from  a $U(4)$ symmetry
realized dynamically in terms of 3 subgroup chains.  One of those subgroups is
the $SO(5)$ subgroup discussed extensively by S. C. Zhang 
\cite{zha97}.
In this paper,
we have used $SU(4)$ coherent states to analyze the
$SO(5)$ energy surface.  We interpret the
Zhang $SO(5)$
as a critical dynamical symmetry that
interpolates between AF and D-wave SC order as
doping is varied, and have noted
similarities with analogous critical dynamical symmetries from
nuclear structure physics.  This permits the $SO(5)$
symmetry to be understood dynamically as a critical phase that for a
range of doping has an energy surface extremely soft
against AF fluctuations and therefore having much of the
character of a spin glass.  
Thus, we propose that slightly broken $SO(5)$ is the symmetry of the
underdoped regime, but that AF and optimally doped SC compounds are described
by different subgroups of $SU(4)$.
 
L.-A. Wu was supported in part by the National Natural Science Foundation
of China.
Research at
Chung Yuan Christian University is supported by the National Science
Council of the ROC
through contracts  {\it NO.} NSC 88-2112-M033-006.

\baselineskip = 14pt
\bibliographystyle{unsrt}

\begin{figure}
\caption{Coherent state 
energy surfaces.  The energy unit is 
$G_0\Omega^2 / 4$ (see Eq.\ (\protect\ref{eq20})). 
Numbers on curves are twice the
lattice occupation fractions, 
with $n / \Omega = 1$ corresponding to half filling and $0 < n / \Omega < 1$
to finite hole doping.
$SO(5)$ symmetry corresponds to $p=1/2$.}
\label{fig1}
\end{figure}

\begin{figure}
\caption{(a)~As for Fig. 1, but for slightly perturbed $SO(5)$.
The dotted line indicates the classical ground
state as $n$ varies.  (b)~Variation of the
AF order parameter with occupation number for different values of $p$.}
\label{fig2}
\end{figure}

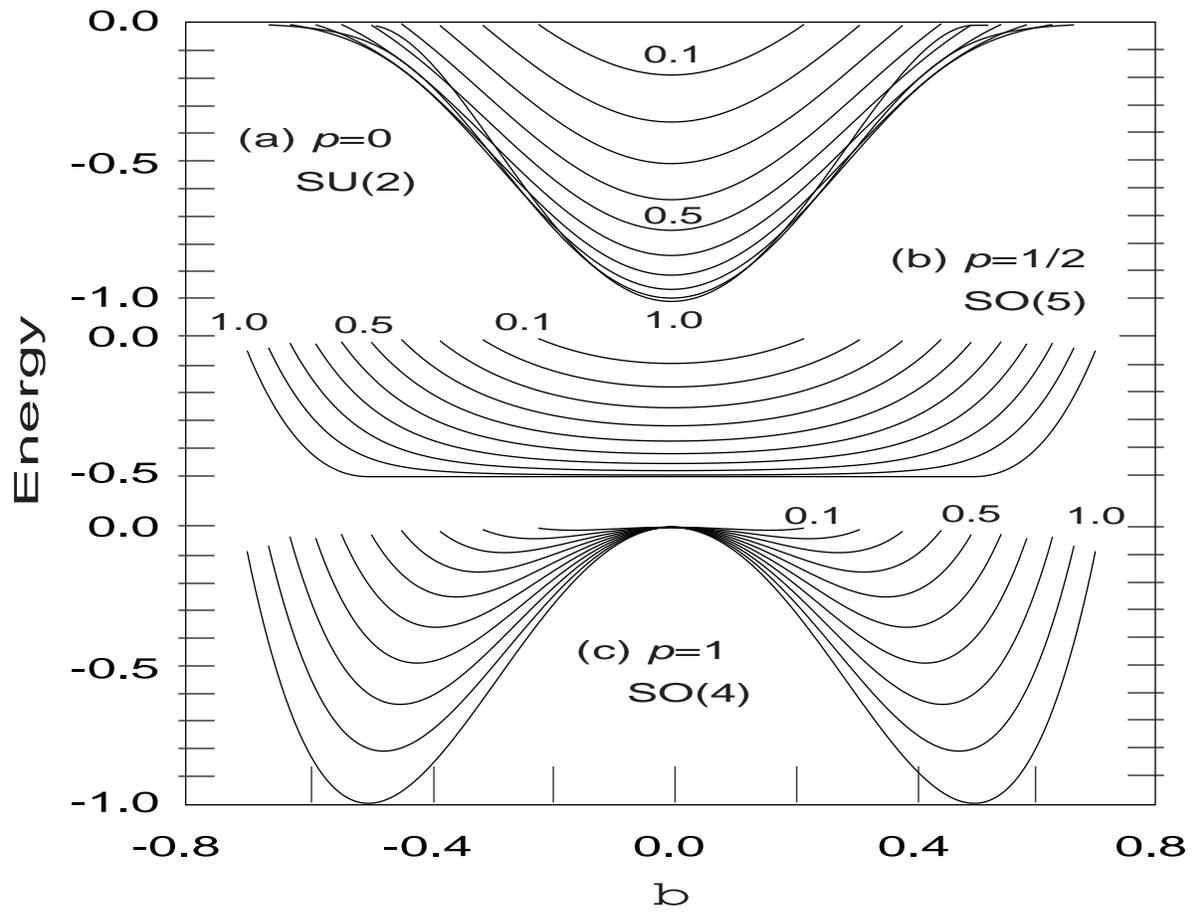
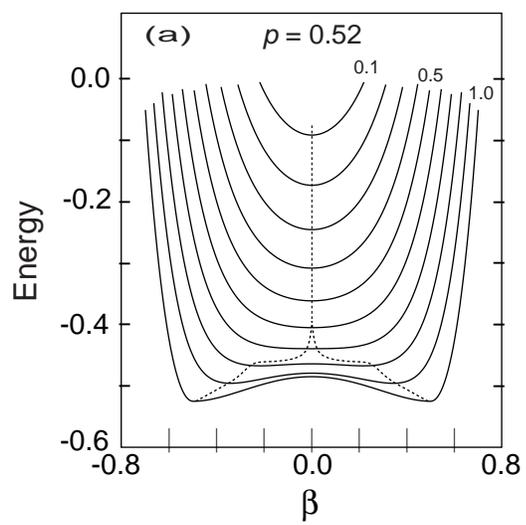
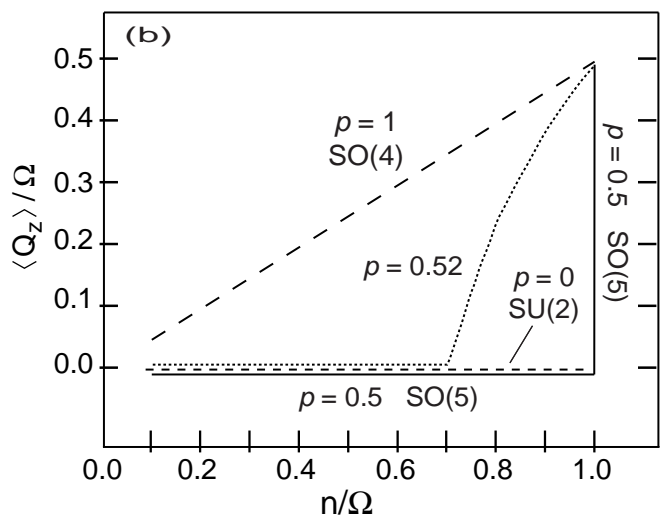
\end{document}